\documentclass[%
10pt,
twocolumn,
oneside,
superscriptaddress,
nofootinbib,
amsmath,
amssymb,
aps,
]{revtex4-2}

\usepackage{graphicx}
\usepackage{dcolumn}
\usepackage{bm}
\usepackage{xcolor}

\makeatletter
\def\tagform@#1{(\textcolor{magenta}{#1})}
\makeatother

\makeatletter
\renewcommand{\eqref}[1]{\textup{\tagform@{\textcolor{magenta}{\ref*{#1}}}}}
\makeatother

\usepackage[colorlinks=true,linkcolor=magenta,citecolor=magenta,urlcolor=magenta]{hyperref}
\usepackage{tikz-feynman}
\usepackage{nicefrac}
\usepackage{lipsum}

\begin{document}

\title{\color{teal}\bf Rethinking Dimensional Regularization in Critical Phenomena}

\author{P. Beretta}
\email{pberetta@fing.edu.uy}
\affiliation{IFFI, Universidad de la Rep\'ublica, J.H.y Reissig 565, 11300 Montevideo, Uruguay}

\author{A. Codello}
\email{alessandro.codello@unive.it}
\affiliation{DMSN, Ca'\ Foscari University of Venice, Via Torino 155, 30172 - Venice, Italy}
\affiliation{IFFI, Universidad de la Rep\'ublica, J.H.y Reissig 565, 11300 Montevideo, Uruguay}

\date{\today}

\begin{abstract}
We show that it is possible to use dimensional regularization (DR) beyond the usual $\varepsilon$-expansion in the context of renormalization group (RG) calculations in Critical Phenomena. 
Based on this fact, we propose a new functional RG scheme -- {\it Functional Dimensional Regularization} (FDR) -- and apply it to a scalar theory in three dimensions.
We compute the critical exponents of the {\tt Ising} universality class directly in $d=3$ under various typical approximations.
The method that emerges combines the ``agility" typical of DR with the ``generality" proper of functional RG.
Moreover, at a given order of approximation, FDR seems to provide faster convergence and better estimates than other functional RGs.
\end{abstract}

\maketitle

\subsection*{\color{teal}Introduction}\vspace{-0.2cm}

Since the early days of the field theory approach to critical phenomena,
dimensional regularization \cite{Bollini:1972ui,tHooft:1972tcz,tHooft:1973mfk} (DR) has been used as the main tool to perform renormalization group (RG) calculations in the context of the $\varepsilon$-expansion \cite{Wilson:1971dc,wilson1972feynman,Wilson:1973jj}, where one studies a fixed point theory just below its upper critical dimension $d_c$, in a power series in the variable $\varepsilon = d_c-d$.
Indeed, a vast amount of work has been done over decades -- and is still done today -- using these two techniques combined, making DR and the $\varepsilon$-expansion the standard, preferred, approach used in the field of Critical Phenomena. It was only in later stages that researchers shifted to the use of exact functional RG equations \cite{Polchinski:1983gv,Wetterich:1989xg,Morris:1993qb} and other functional RG frameworks \cite{Wegner:1972ih,Liao:1994fp}, mainly to overcome the limitations of the $\varepsilon$-expansion in which the study of RG properties is reliable -- in absence of re-summations \cite{Zinn-Justin:2002ecy} -- only near the upper critical dimension.
These functional RGs are all based on an explicit mass cutoffs, i.e. a regulator which is either ultraviolet (UV) or infrared (IR) -- and the widespread opinion is that the price to pay to have an RG formalism able to work in arbitrary dimension, in particular well below $d_c$, is the hurdle of dealing with all the additional scheme dependences and complications that a mass regulator introduces in comparison to the ``agility" 
typical of DR.

In this Letter we show that it is instead possible to construct a functional RG -- i.e. a framework that is valid in arbitrary dimension and that probes theory space at a functional level -- based entirely on DR  without the need to introduce an explicit mass cutoff -- thus avoiding all related scheme dependencies and complications. The main insight -- that has been completely missed to date -- is that in the renormalization procedure {\it all DR poles in the complex $d$-plane have to be subtracted}, not only those associated with the particular critical dimension under study  (dimension, we recall, that is normally chosen on top of a given interaction, for example $\lambda \phi^4$ is associated to $d_c=4$ and vice-versa). 
Here we show that the inclusion of all poles/critical dimensions -- or equivalently all scheme independent contributions to the RG beta functions -- leads to a functional flow with all the desired properties: applicability in general dimension, threshold behaviour, scaling solutions description of fixed points, well posed eigen-perturbations and  so on. Moreover, we discover that our approach -- that we will call {\it Functional Dimensional Regularization} (FDR) -- displays fast convergence and furnishes particularly good estimates if compared to other functional RGs at the same level of approximation. Our insight opens a wide new spectrum of possibilities and is the starting point for the construction of a functional RG that minimizes scheme dependence, by eliminating altogether the need of an explicit mass cutoff.

We start this Letter by introducing the FDR approach in a pedagogical way, by revealing the underlying pattern present in DR beta functions $\beta_i^{\rm DR}$ (which encode the RG flow of the coupling constants) of the usual scalar theory, when analyzed in parallel at all different critical dimensions that appear at one-loop order. Following this pattern we are led to reconstruct the exponential threshold functions proper of DR. This picture then naturally suggests our main insight:  {\it FDR beta functions are the sum of DR beta functions over all critical dimensions}. Then we show that this picture can be naturally re-casted in the normal formulation of DR if we indeed subtract all poles -- as was hinted by S. Weinberg in a note appearing in \cite{Weinberg} -- and not just the one related to the critical dimension of interest. The functional generalization at this point is straightforward. As a first application we study the Wilson-Fisher fixed point in three dimensions -- describing the {\tt Ising} universality class -- to find very encouraging numerical estimates for the critical exponents, in particular for the anomalous dimension.

\subsection*{\color{teal}A pattern in the beta functions}\vspace{-0.2cm}
\begin{table*}[t!]
\centering
\begin{tabular}{cccccccccc}
$\mu^{d-2}$ & & $\mu^{d-4}$ & & $\mu^{d-6}$ & & $\mu^{d-8}$ & & & \\
$-\lambda_{4}$ 
& & $+\lambda_{2}\lambda_{4}$ 
& & $-\frac{1}{2}\lambda_{2}^{2}\lambda_{4}$ 
& & $\cdots$ 
& $\to$ & 
$\mu^{d-2}(-\lambda_{4})e^{-\frac{\lambda_{2}}{\mu^2}} $ & $ \equiv \beta_2^{\rm FDR}$ \\
$-\lambda_{6}$ 
& & $+3\lambda_{4}^{2}+\lambda_{2}\lambda_{6}$ 
& & $-\frac{1}{2}\lambda_{6}\lambda_{2}^{2}-3\lambda_{4}^{2}\lambda_{2}$ 
& & $\cdots$ 
& $\to$ & 
$\mu^{d-4}(3\lambda_{4}^{2}-\mu^{2}\lambda_{6})e^{-\frac{\lambda_{2}}{\mu^2}}  
$&$ \equiv \beta_4^{\rm FDR}$ \\
$-\lambda_{8}$ 
& & $+15\lambda_{4}\lambda_{6}+\lambda_{2}\lambda_{8}$ 
& & $-15\lambda_{4}^{3}-15\lambda_{2}\lambda_{6}\lambda_{4}-\frac{1}{2}\lambda_{2}^{2}\lambda_{8}$ 
& & $\cdots$ 
& $\to$ & 
$\mu^{d-6}(-15\lambda_{4}^{3}+15\mu^{2}\lambda_{4}\lambda_{6}-\mu^{4}\lambda_{8})e^{-\frac{\lambda_{2}}{\mu^2}}  $&$\equiv \beta_6^{\rm FDR}$ \\
$\vdots$ & & $\vdots$ & & $\vdots$ & & $\vdots$ & & $\vdots$ & $\vdots$
\end{tabular}
\caption{The table displays the DR beta functions of Eqs. \eqref{betas4}, \eqref{betas6}, \eqref{betas2} organized by their extra dimensionality factor $\mu^{d-d_c}$. The passage from DR to FDR is obtained by summing each line, thus defining the FDR beta functions $\beta_i^{\rm FDR}$ as the {\it sum over all upper critical dimensions} of DR beta functions $\beta_i^{\rm DR}$, each properly multiplied by $\mu^{d-d_c}$ to ensure the correct dimensionality.}
\label{betas}
\end{table*}

In perturbation theory -- when we adopt DR -- divergences can be computed from 't Hooft formula \cite{Vassilevich:2003xt} or using ordinary Feynman diagrams \cite{Weinberg:1995mt,Zinn-Justin:2002ecy}.
Consider the $\lambda \phi^{4}$ theory which is normally analyzed in $d_c=4$ where the coupling $\lambda_4 \equiv \lambda $ is marginal.
One finds the usual scheme independent beta function $\beta_{4}^{\rm DR}=\frac{3\lambda^{2}}{(4\pi)^{2}}$.
In presence of a mass term $\lambda_2 \equiv m^{2}$ there is an associated beta function generated by the field self-interaction 
$\beta_{2}^{\rm DR}  =\frac{\lambda_{2}\lambda_{4}}{(4\pi)^{2}}$.
If we also consider couplings that are irrelevant in four dimensions --- $\lambda_6, \lambda_8, \lambda_{10}, ...$ -- we would additionally find the following tower of beta functions:\footnote{We scale away $4\pi$'s for aid of reading from now on.}
\begin{eqnarray}\label{betas4}
\beta_{2}^{\rm DR}(d_c\!=\!4) & = & \lambda_{2}\lambda_{4} \nonumber\\
\beta_{4}^{\rm DR}(d_c\!=\!4) & = & 3\lambda_{4}^{2}+\lambda_{2}\lambda_{6}\nonumber\\
\beta_{6}^{\rm DR}(d_c\!=\!4) & = & 15\lambda_{4}\lambda_{6}+\lambda_{2}\lambda_{8}\nonumber\\
\beta_{8}^{\rm DR}(d_c\!=\!4) & = & 35\lambda_{6}^{2}+28\lambda_{4}\lambda_{8}+\lambda_{2}\lambda_{10}
\nonumber\\&\vdots&
\end{eqnarray}
Note that we are dealing with dimension-full couplings and that only $\lambda_4$ equals its dimension-less counterpart in four dimensions.
What happens if instead we renormalize the theory in $d_c=6$?
Here we will be naturally interested in the $\lambda_{3}\phi^{3}$ coupling, which is marginal in six dimensions,
but it is a fact that also the previous ``even" couplings do run and have scheme independent beta functions in DR at $d_c=6$.
In particular, we would find:\footnote{Around $d_c=6$ or as before around $d_c=4$, we could easily add all the ``odd'' couplings $\lambda_1, \lambda_3, \lambda_5, ...$ but to avoid useless confusion we leave them out for the moment. All these couplings will be automatically included in the following in the functional formulation.}
\begin{eqnarray}\label{betas6}
\beta_{2}^{\rm DR}(d_c\!=\!6) & = & -\tfrac{1}{2}\lambda_{2}^{2}\lambda_{4}\nonumber\\
\beta_{4}^{\rm DR}(d_c\!=\!6) & = & -\tfrac{1}{2}\lambda_{6}\lambda_{2}^{2}-3\lambda_{4}^{2}\lambda_{2}\nonumber\\
\beta_{6}^{\rm DR}(d_c\!=\!6) & = & -15\lambda_{4}^{3}-15\lambda_{2}\lambda_{6}\lambda_{4}-\tfrac{1}{2}\lambda_{2}^{2}\lambda_{8}\nonumber\\
\beta_{8}^{\rm DR}(d_c\!=\!6) & = & -\tfrac{1}{2}\lambda_{10}\lambda_{2}^{2}-35\lambda_{6}^{2}\lambda_{2}-28\lambda_{4}\lambda_{8}\lambda_{2}
\nonumber\\&&
-210\lambda_{4}^{2}\lambda_{6}
\nonumber\\&\vdots&
\end{eqnarray}
Let's also look at renormalization in $d_c=2$ where we would easily obtain the following DR beta functions for our set of couplings:
\begin{eqnarray}\label{betas2}
\beta_{2}^{\rm DR}(d_c\!=\!2) & = &-\lambda_{4}\nonumber\\
\beta_{4}^{\rm DR}(d_c\!=\!2) & = &-\lambda_{6}\nonumber\\
\beta_{6}^{\rm DR}(d_c\!=\!2) & = &-\lambda_{8}\nonumber\\
\beta_{8}^{\rm DR}(d_c\!=\!2) & = &-\lambda_{10}
\nonumber\\&\vdots&
\end{eqnarray}
The beta functions \eqref{betas4}, \eqref{betas6} and \eqref{betas2} are all dimension-full and thus have a specific mass dimension at those $d_c$'s where they have been computed.
If we move away from these $d_c$'s they will acquire an extra dimensionality factor of the form $\mu^{d-d_c}$, with $\mu$ the DR mass scale.
To spot the pattern that we are after let us arrange the beta functions as in Tab.~\ref{betas}, organized by their extra dimensionality factor -- keeping in mind that the beta functions in \eqref{betas4} must be multiplied by $\mu^{d-4}$, those in \eqref{betas6} by $\mu^{d-6}$ and those in \eqref{betas2} by $\mu^{d-2}$.
Now the pattern is apparent: an exponential threshold function 
builds up if we sum along each row and if we add the beta functions terms relative to all the other one-loop critical dimensions $d_c = 8,10,12,...$.
The expressions emerging from this procedure define our FDR beta functions. 
As we will see in the following of this Letter, the RG flow constructed in this way has all the characteristics that are normally associated with {\it functional} RG flows -- from here our name {\it functional} DR.

Equivalently, it is clear that what we are doing is defining a scheme (FDR) where the beta functions are build combining scheme independent DR beta functions from all $d_c$'s.
More precisely, we use scheme freedom to assign the structure $\beta^{\rm DR}_i(d_c)$ -- i.e. the scheme independent logarithmic divergence coefficient -- to the power divergence $\mu^{d-d_c}$ (which is IR or UV depending on the value of $d$). Here $d$ is crucially free to take any positive value, thus forcing the summation to cover {\it all} critical dimensions.
This leads to our master formula at one-loop order that defines FDR beta functions as a sum over $d_c$ of the respective DR beta functions, multiplied by their extra dimensionality factor:
\begin{equation}\label{MasterFormulaCouplings}
 \beta_i^{\rm FDR}(d)=\sum_{d_{c}}\mu^{d-d_{c}}\beta_i^{\rm DR}(d_{c})  \,.
\end{equation}
The sum \eqref{MasterFormulaCouplings} includes all one-loop critical dimensions $d_{c}=2,4,6,8,10,...$.
Since the DR beta functions $\beta_i^{\rm DR}(d_{c})$ are directly related to the OPE coefficients of the corresponding Gaussian theory in $d_c$, one can imagine computing the r.h.s. of \eqref{MasterFormulaCouplings} using CFT \cite{Rychkov:2015naa,Gopakumar:2016wkt,Codello:2018nbe} or other methods, since the information it contains is universal.
Finally, the crucial point we need to remark -- and which is the main difference with respect to previous proposals -- is that IR dimensions (those that diverge when $\mu \to 0$ at a given $d$) are fundamental to the definition of the FDR flow since they allow $d$ to vary freely and additionally build up into the exponential threshold functions emerging in Tab.~\ref{betas}.

\subsection*{\color{teal}...and might turn out to be useful \cite{Weinberg}}\vspace{-0.2cm}

What we did in the previous section can be accommodated in the standard DR work-flow if we consistently subtract {\it all} poles in the complex $d$-plane and not only the one at $d_c=4$, as usually done when studying the {\tt Ising} universality class within the $\varepsilon$-expansion\footnote{Or similarly other specific critical dimensions, like $d_c=3$ for the {\tt Tricritical} class and $d_c=6$ in the case of {\tt LeeYang}.}.
In \cite{Weinberg} Weinberg hints at a similar procedure, but does not follow further the intuition -- and in any case, what he had in mind was the summation uniquely over (finitely many) UV dimensions, and not over UV and (infinitely many) IR dimensions, as we are proposing in this Letter. Extending the seminal paper of 't Hooft  \cite{tHooft:1973mfk}, he gives the following formula for dimension-less beta functions:
\begin{equation}\label{Weinberg}
\tilde \beta_{i}(d)=-d_{i}\tilde\lambda_{i}+\sum_{d_{c}}\Big\{ -\rho_{i}B_{1}^{i}(d_{c})+\sum_j\partial_{j}B_{1}^{i}(d_{c})\rho_{j}\tilde\lambda_{j}\Big\}\,,
\end{equation}
where $\tilde\lambda_{i} = \mu^{-d_i}\lambda_{i}$ are the dimension-less couplings, with $d_{i}=\sigma_{i}+d\rho_{i}$, $\rho_{i}=1-i/2$ and $\sigma_{i}=i$.
Here $B_{1}^{i}(d_{c})$ are the residue functions at the simple poles $\tfrac{1}{d-d_{c}}$.
For the scalar action that we are considering,
\begin{equation}
S=\int {\rm d}^dx\left\{ \frac{1}{2}(\partial\phi)^{2}+\frac{\lambda_{2}}{2!}\phi^{2}+\frac{\lambda_{4}}{4!}\phi^{4}+\frac{\lambda_{6}}{6!}\phi^{6}+...\right\}\,,  \label{action}
\end{equation}
the one-loop counter terms are $\Delta S(d_c) =-\Gamma|_{\infty}$ and can be computed using 't Hooft  formula \cite{Vassilevich:2003xt}:
\begin{equation}
\Delta S(d_c) =\frac{1}{d_c-d} {\mathcal B}_{d_{c}}(-\partial^{2}+V'')\,,
\end{equation}
where the Heat Kernel coefficients are just ${\mathcal B}_{d}(-\partial^{2}+V'') = \tfrac{1}{(d/2)!} \int(-V'')^{d/2}$ in this simple scalar example.
When the residues functions are written in terms of the dimension-less couplings, a simple computation gives:
\begin{eqnarray*}
B_{1}^{2}(d_{c}) & =&\tfrac{1}{(\tfrac{d_{c}}{2})!}\tfrac{d_{c}}{2}\tilde\lambda_{4} (-\tilde\lambda_{2}){}^{\frac{d_{c}-2}{2}}\\
B_{1}^{4}(d_{c}) & =&-\tfrac{1}{(\tfrac{d_{c}}{2})!}\tfrac{d_{c}}{4}\left[3(d_{c}-2)\tilde\lambda_{4}^{2}+2\tilde\lambda_{2}\tilde\lambda_{6}\right] (-\tilde\lambda_{2}){}^{\frac{d_{c}-4}{2}}\\
B_{1}^{6}(d_{c}) & =&\tfrac{1}{(\tfrac{d_{c}}{2})!}\tfrac{d_{c}}{8}\big[15(d_{c}-2)(d_{c}-4)\tilde\lambda_{4}^{3} \\\ &&+\,30(d_{c}-2)\tilde\lambda_{2}\tilde\lambda_{4}\tilde\lambda_{6}+4\tilde\lambda_{2}^{2}\tilde\lambda_{8}\big](-\tilde\lambda_{2}){}^{\frac{d_{c}-6}{2}}
\end{eqnarray*}
and so on.
Now we can compute the beta functions using Weinberg's formula \eqref{Weinberg}, but summing over {\it all} $d_c$'s.
We obtain exactly the dimension-less version fo the beta functions\footnote{In this and later sections we drop the super-script FDR on beta functions/functionals.} of Tab.~\ref{betas}:
\begin{eqnarray}
\tilde\beta_{2} & = & (-2\!+\!\eta)\tilde\lambda_{2}\!-\!\tilde\lambda_{4} e^{-\tilde\lambda_{2}}\nonumber\\
\tilde\beta_{4} & = & (d\!-\!4\!+\!2\eta)\tilde\lambda_{4}\!+\!(3\tilde\lambda_{4}^{2}\!-\!\tilde\lambda_{6})e^{-\tilde\lambda_{2}}\nonumber\\
\tilde\beta_{6} & = & (2d\!-\!6\!+\!3\eta)\tilde\lambda_{6}\!-\!(15\tilde\lambda_{4}^{3}\!-\!15\tilde\lambda_{6}\tilde\lambda_{4}
\!+\!\tilde\lambda_{8}) e^{-\tilde\lambda_{2}}
\nonumber\\& \vdots &
\label{betadimless}
\end{eqnarray}
%
where we have included the anomalous dimension $\eta$ for later use. 
To conclude this section, we stress again that it is the sum over both UV and IR critical dimensions the key crucial difference with respect to  Weinberg's {\it ...and might turn out to be useful} and also with respect to other proposals that appeared over the years -- see \cite{Jack:1990pz,Kaplan:1998tg,Phillips:1998uy,Kluth:2024lar} just to mention some, that used UV logarithmic divergencies in lower dimensions to mimic positive mass power contributions to the beta functions in $d=4$. 
The application to the Wilson-Fisher fixed point, that we present in later sections, serves just as an example of the fundamental effect that the inclusion of all IR divergences has on the quality of the RG flow and precision of critical exponents estimates.

\subsection*{\color{teal}Towards a functional flow in DR}\vspace{-0.2cm}

The generalization to a functional flow is now an easy task. All DR beta functions discussed so far are more efficiently derived by a simple functional generator in each critical dimension. For example, the $d_{c}=4$ beta functions are generated by 
$
\beta_{V}^{\rm DR}(\phi)|_{d_c=4}=\tfrac{1}{2}(V'')^{2}
$
in which we insert $\beta^{\rm DR}_{V}(\phi)=\frac{1}{2}\beta_{2}^{\rm DR}\phi^{2}+...$
on the l.h.s. and $V(\phi)=\frac{1}{2}\lambda_{2}\phi^{2}+...$ on the r.h.s. and series expand to match powers. The advantage of working functionally, even at the standard perturbative level, is manifold: simpler calculations condensed in the determination of few universal constants (universal data), access to some families of OPE coefficients, straightforward generalization to general $N$ --
and are discussed at length in \cite{ODwyer:2007brp,Codello:2017hhh,Zinati:2019gct,BenAliZinati:2021rqc}.
These papers also give the functionals $\beta_{V}^{\rm DR}(\phi)|_{d_c =2}=-V''$ and  $\beta_{V}^{\rm DR}|_{d_c =6}=\tfrac{1}{6}(-V'' )^{3}$.
The general expression is not difficult to guess \cite{FDR}:
\begin{equation}\label{betaVDR}
\beta_{V}^{\rm DR}(d_{c})=\frac{1}{(\frac{d_{c}}{2})!} (-V'' )^{\frac{d_{c}}{2}}\,.
\end{equation}
The master formula for the couplings \eqref{MasterFormulaCouplings} straightforwardly generalises to {\it functionals}:
\begin{equation}\label{masterL}
\beta_V(d)=\sum_{d_{c}}\mu^{d-d_{c}}\beta^{\rm DR}_V(d_{c})\,.
\end{equation}
Using now equation \eqref{betaVDR} we immediately find
\begin{equation}\label{betaV}
\beta_{V}(d)=\sum_{d_{c}}\mu^{d-d_{c}}\frac{1}{(\frac{d_{c}}{2})!}(-V'' )^{\frac{d_{c}}{2}}=\mu^{d}e^{-\frac{V''}{\mu^2}}\,,
\end{equation}
were we summed over $d_c=0,2,4,6,...$.
This is the FDR flow for the effective potential at one-loop order. It generates all dimension-full beta functions, the first of which are displayed in Tab.~\ref{betas}.
The RG flow \eqref{betaV} can now be used in any dimension, we will apply it to the Wilson-Fisher fixed point in three dimensions, which describes the {\tt Ising} universality class.
Finally, we note that the flow \eqref{betaV} is  exactly the same obtained in the framework of the proper-time RG (PTRG) in the limit $m \to \infty$ and after a field redefinition \cite{Bonanno:2000yp,Mazza:2001bp,Zappala:2002nx}. 
We will comment more on this connection in a later section.

\begin{figure}
\centering
\includegraphics[width=0.45\textwidth]{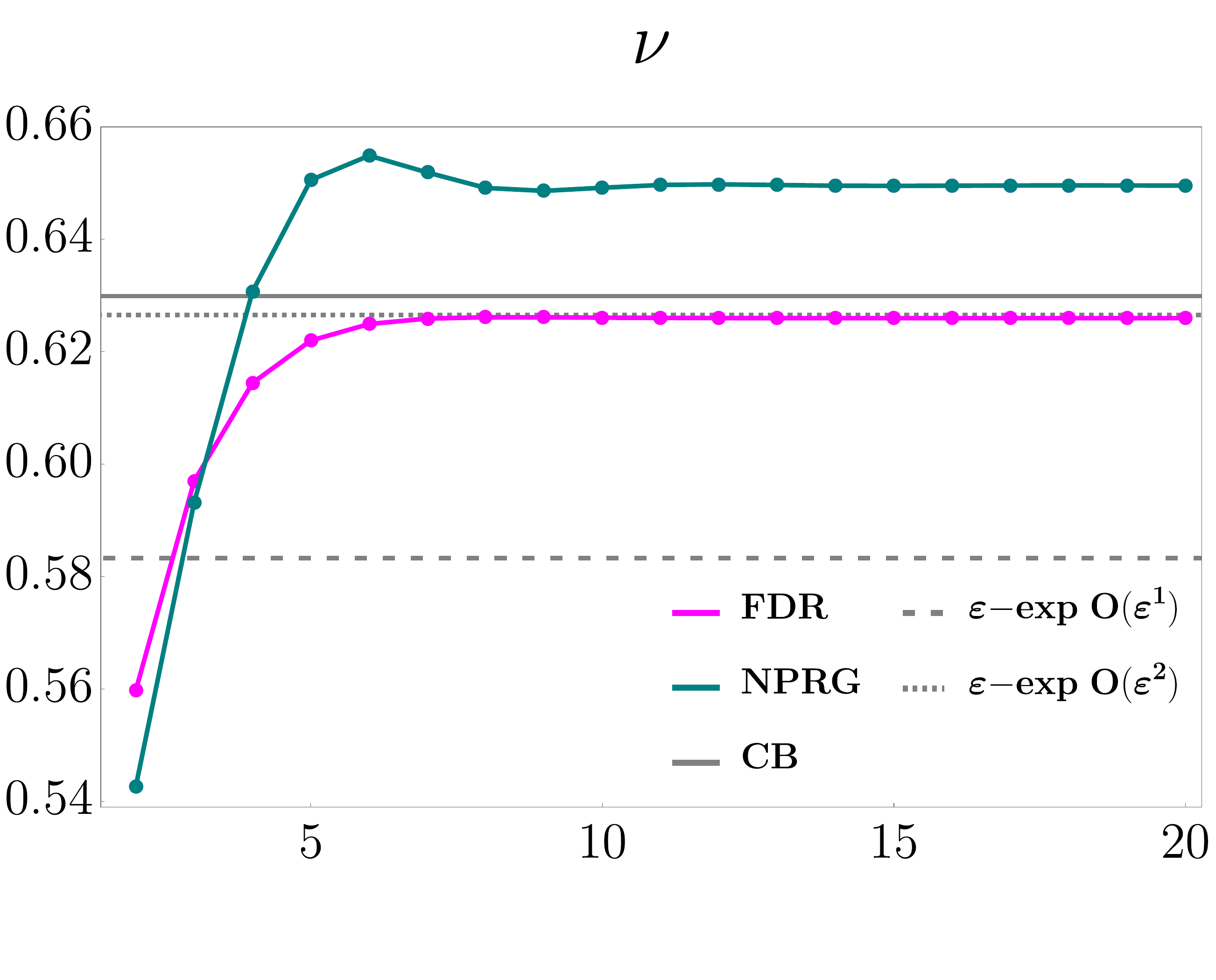}\vspace{-0.2cm}
\includegraphics[width=0.45\textwidth]{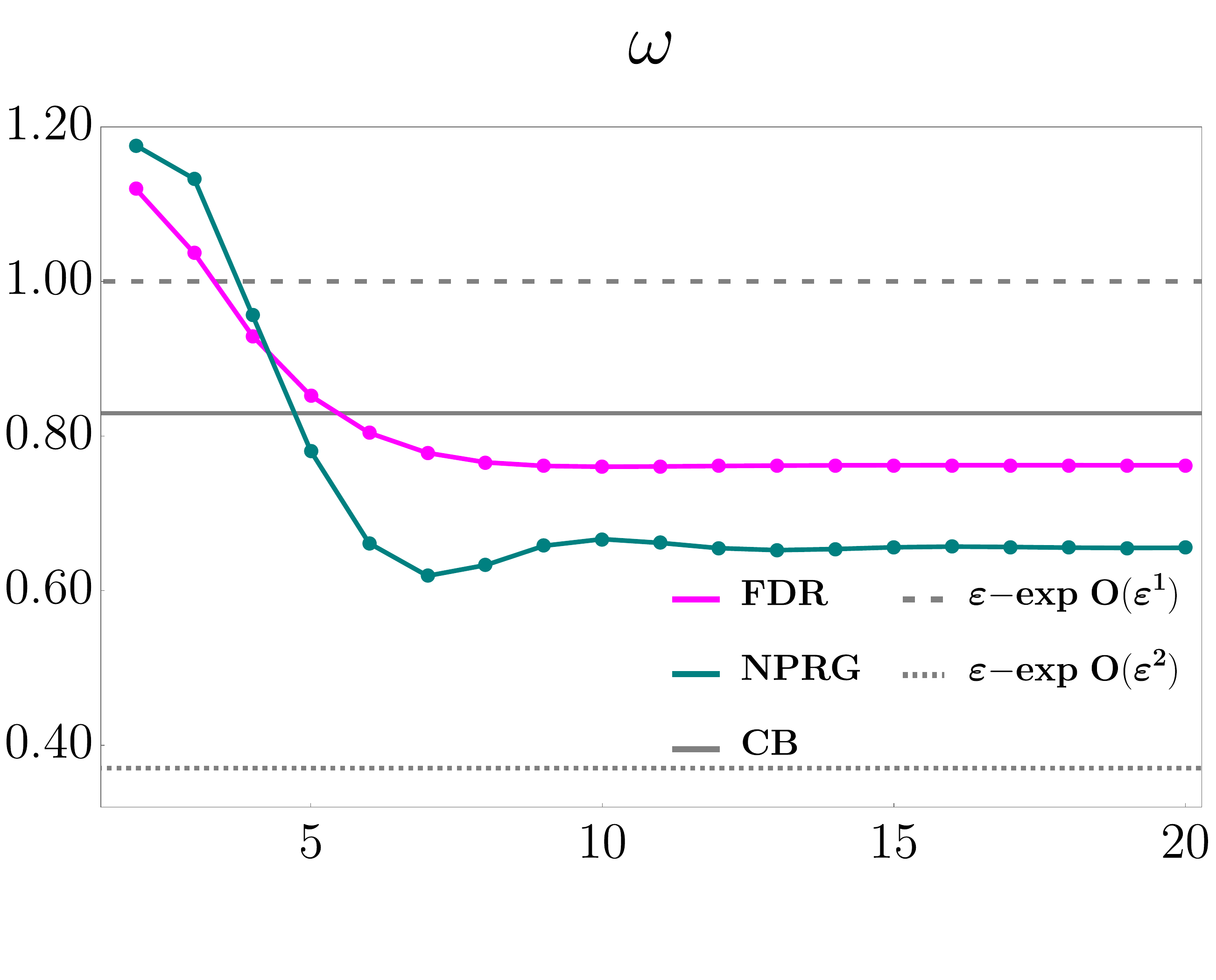}\vspace{-0.5cm}
\caption{Convergence in successive truncations of the critical exponents $\nu$ and $\omega$ in $d=3$. Comparison is shown among the FDR-LPA (magenta) and NPRG-LPA (teal). We also report $\varepsilon$-expansion and CB values for comparison.}
\label{NuOmegaFDRLPA}
\end{figure}

\subsection*{\color{teal}Wilson-Fisher fixed point in $d=3$}\vspace{-0.2cm}

We now focus on the main problem our formalism is build to address: the computation of critical exponents directly in three dimensions, where the only single field unitary universality class is {\tt Ising}.
Before deriving the beta functions from the beta functionals, we need to rewrite these in terms of dimension-less variables,
$\tilde V(\varphi)\equiv  \mu^{-d} V(\varphi \mu^{\frac{1}{2}(d-2+\eta)})$,
where $\varphi$ the dimension-less field.
The dimension-less beta functional $\beta_{\tilde V} \equiv \mu\, \partial_\mu \tilde V$ is easily obtained from \eqref{betaV} and takes the simple form
\begin{equation}\label{dimlessbetav}
\beta_{\tilde V}  = -d \,\tilde V+\tfrac{d-2+\eta}{2} \varphi\, \tilde V'  + e^{-\tilde V''}\,.
\end{equation}
If we expand the dimension-less potential in Taylor series,
then the coefficients are the dimension-less couplings $\tilde\lambda_i$.
Expecting $\mathbb{Z}_2$ symmetry at the fixed point we keep only even couplings,
$\tilde V(\varphi)=\sum_{i=1}^{N_{tr}}\tfrac{1}{(2i)!}\tilde\lambda_{2i}\varphi^{2i}$,
where $N_{tr}$ is the truncation order.
The dimension-less beta functions $\tilde\beta_i \equiv \mu \partial_\mu \tilde\lambda_i$ are straightforwardly extracted inserting the Taylor expansion into \eqref{dimlessbetav} and then comparing equal powers of the field.
To determine the existence of a universality class in a given $d$ we solve for the fixed points
$\tilde\beta_i = 0$ and check their stability for increasing truncation order (i.e. larger and larger values of $N_{tr}$).
Critical exponents are then computed as eigenvalues of the stability matrix $M_{ij}=\partial \tilde\beta_i/\partial\tilde\lambda_j|_*$.

We now consider the Local Potential Approximation (LPA) where the only running couplings are those of the potential and the anomalous dimension is set to zero.
For example, at $N_{tr}=2$ the FDR-LPA beta functions are the first two from Eq. \eqref{betadimless} with $\tilde \lambda_6 = 0$ on the r.h.s., while at  $N_{tr}=3$ we keep all three equations but set  $\tilde \lambda_8 = 0$. The system of equations for higher $N_{tr}$ is similarly determined.
The results of the full analysis are shown in Fig.~\ref{NuOmegaFDRLPA} for the critical exponents $\nu$ and $\omega$.
The fact that our estimates are close to Conformal Bootstrap (CB) values \cite{Chang:2024whx} is quite remarkable considering the simplicity of the approximation involved: a one-loop computation using DR with the only twist of summing over all critical dimensions.
It is also evident that there is an improved convergence with respect to the traditional LPA of the non-perturbative RG (NPRG). 
We are comparing with the optimized Litim cutoff,  which in turn is spectrally equivalent to the scheme independent Polchinski LPA \cite{Morris:2005ck,Bervillier:2007rc,Litim:2005us}. Thus at LPA level, FDR is better than both optimized NPRG and Polchinski (and equivalent to PTRG).
We will extend the fixed point analysis to scaling solutions and eigen-perturbations in our companion paper \cite{FDR}.

\subsection*{\color{teal}Anomalous dimension}\vspace{-0.2cm}

\begin{figure}
\centering
\includegraphics[width=0.45\textwidth]{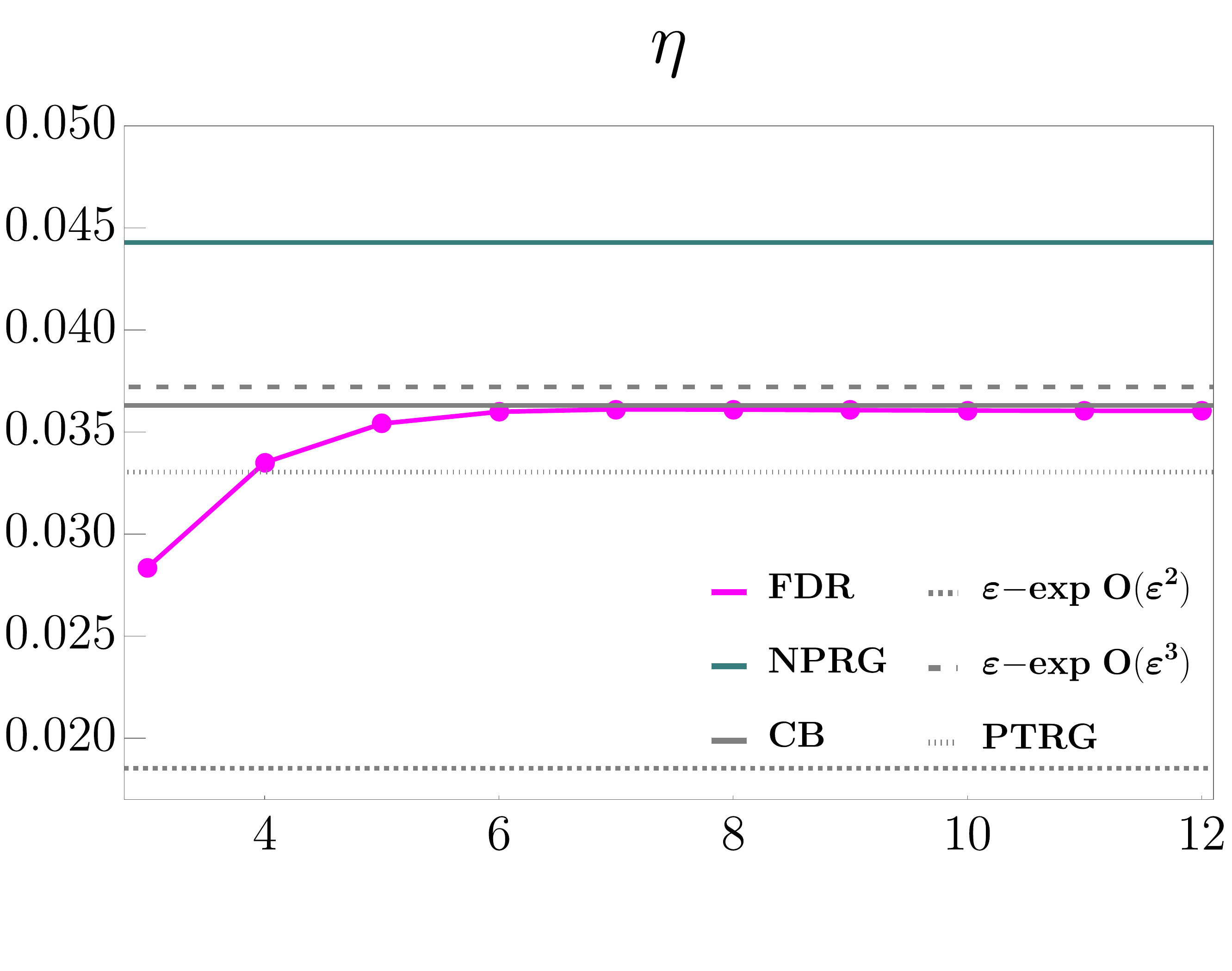}\vspace{-0.5cm}
\caption{
Convergence of the critical exponent $\eta$ in the FDR approximation discussed in the text (magenta) in $d=3$.
We also display reference results from state-of-the-art CB \cite{Chang:2024whx}; second order of the derivative expansion of the NPRG (teal), with the $\Theta^2$ cutoff \cite{DePolsi:2020pjk}; and the $m\to\infty$ limit of PTRG \cite{Bonanno:2000yp}.
Dashed lines report $\varepsilon$-expansion $O(\varepsilon^2)$ and $O(\varepsilon^3)$ results (without re-summation).}
\label{EtaFDR}
\end{figure}

To make any significant study of critical properties we need to include the anomalous dimension $\eta$.
One has to compute the flow of the field dependent wave-function-renormalization $Z$ induced by the potential $V$. 
Then $\eta$ is determined self-consistently by imposing the normalization condition $\tilde Z(0)=0$ on the dimension-less flow
\begin{equation}
\beta_{\tilde Z} =\eta \, (1+\tilde Z) +\tfrac{d-2+\eta}{2}\varphi\, \tilde Z' + \mu^{\eta}\beta_Z\,,
\end{equation}
where $\tilde Z(\varphi)\equiv\mu^{\eta}Z(\varphi\mu^{\frac{1}{2}(d-2+\eta)})$.
Setting $\beta_{\tilde Z}(0)=0$ directly gives
$\eta = -\mu^{\eta}\beta_Z |_{\varphi \to 0}$.
A calculation similar to the one of $\beta_V$ of the previous section (details will be reported in \cite{FDR}) gives the following functional flow
\begin{equation}\label{betaZL1}
\beta_Z(d) = -  \mu^{d-6} \,\tfrac{1}{6} (V''')^2 \,e^{-\frac{V''}{\mu^2}}\,.
\end{equation}
From this result we immediately deduce $\eta = \tfrac{1}{6}\tilde\lambda_3^2 \,e^{-\tilde\lambda_2}$.
This contribution clearly vanishes for an ``even" theory like {\tt Ising} -- it would be the leading contribution in the case of an ``odd" theory like {\tt Lee-Yang}. 
In the even case, the leading contribution to the anomalous dimension comes from the inclusion of the wave-function-renormalization $Z$ in the r.h.s. of the beta functionals, i.e. we must include it in the action \eqref{action} as a composite operator of the form $\int Z(\phi)\tfrac{1}{2} (\partial \phi)^2$.
A more general calculation readily gives the additional terms \cite{FDR}:
\begin{eqnarray}
\Delta\beta_Z &=&
-\mu^{d-2} Z'' e^{-\frac{V''}{\mu^2}}
+\mu^{d-4} V'''Z' e^{-\frac{V''}{\mu^2}}
\nonumber\\&&
+\mu^{d-6}\tfrac{1}{3} V''V'''Z' e^{-\frac{V''}{\mu^2}}\,. \label{DeltabetaZ}
\end{eqnarray}
We have omitted a higher order contribution proportional to $(Z')^2$ and not included the wave-function-renormalization in the propagator since we are treating it as a composite operator perturbation.
The anomalous dimension correction that follows is
\begin{equation}\label{etaz2}
\Delta\eta = \tilde z_2 \,e^{-\tilde\lambda_2}\,,
\end{equation}
where $\tilde z_2$ is the first coupling appearing in the expansion of the dimension-less wave-function-renormalization functional $\tilde Z(\varphi) = \tfrac{1}{2} \tilde z_2 \varphi^2 + ...\,$.
Equation \eqref{etaz2} is the leading contribution in the case of the {\tt Ising} universality class.
Fig.~\ref{EtaFDR} shows our results for $\eta$ when both the potential and the wave-function renormalization are expanded in Taylor series as described above: as $N_{tr}$ increases ($\tilde Z(\varphi)$ is expanded up to $N_{tr}-1$) convergence to a very good estimate $\eta_{\rm FDR} = 0.03604...$ is achieved with very few couplings ($N_{tr} \sim 12$) at negligible computational cost (order of minutes with {\tt Mathematica} on a laptop).

To conclude this section we note that while $\beta_V$ in \eqref{betaV} is the same as the correspondent PTRG one in the limit $m\to\infty$ \cite{Mazza:2001bp}, the derivative expansion contributions \eqref{DeltabetaZ} crucially differ (for example, PTRG has $d$-dependent coefficients as opposed to the constant ones of FDR); thus the two approaches, albeit having similar beta functionals are not directly related.  One can argue that the ``unreasonable" success of PTRG in the limit $m\to\infty$ is ``explained" by its similarity with the present FDR result which is the functional embodiment of DR. We will present a more detailed comparison between FDR and PTRG in \cite{FDR}.

\subsection*{\color{teal}Two-loop beta function in $d=4$}
\vspace{-0.2cm}

The dimension-less beta function $\tilde{\beta}_{4}$  in \eqref{betadimless} for the marginal coupling $\tilde\lambda_4$ in $d_c=4$
can be expanded to two-loop order considering that $\tilde{\lambda}_{2}\sim\tilde{\lambda}_{4}$
and $\tilde{\lambda}_{6}\sim\tilde{\lambda}_{4}^{3}$ at the Gaussian fixed point 
\begin{equation}
\tilde{\beta}_{4}=\underbrace{3\tilde{\lambda}_{4}^{2}}_{{\rm 1-loop}}+\underbrace{2\tilde{z}_{2}\tilde{\lambda}_{4}-3\tilde{\lambda}_{2}\tilde{\lambda}_{4}^{2}-\tilde{\lambda}_{6}}_{{\rm 2-loop}}+O(\tilde{\lambda}_{4}^{4})\,, \label{beta2loop}
\end{equation}
where furthermore we approximated \eqref{etaz2} as $\eta=\tilde{z}_{2}$ since $\tilde{z}_{2}\sim\tilde{\lambda}_{4}^{2}$. 
Setting  to zero the beta functions of $\tilde{\lambda}_{2}, \tilde{\lambda}_{6}$ and $\tilde{z}_{2}$ one finds
$
\tilde{\lambda}_{2}^{*}=-\tfrac{1}{2}\tilde{\lambda}_{4}$, $\tilde{\lambda}_{6}^{*}
=\tfrac{15}{2}\tilde{\lambda}_{4}^{3}$ and $\tilde{z}_{2}^{*}
=\tfrac{1}{6}\tilde{\lambda}_{4}^{2}
$ 
near the Gaussian fixed point. Inserting now these expressions into \eqref{beta2loop} we recover the two-loop result
for the  beta function of the marginal coupling and for the anomalous dimension \cite{Weinberg:1995mt,Zinn-Justin:2002ecy}:
\begin{eqnarray}\label{betaeta2loop}
\tilde{\beta}_{4}&=&3\tilde{\lambda}_{4}^{2}-\tfrac{17}{3}\tilde{\lambda}_{4}^{3}+O(\tilde{\lambda}_{4}^{4})\nonumber\\
\eta &=& \tfrac{1}{6}\tilde{\lambda}_{4}^2+O(\tilde{\lambda}_{4}^3)\,.
\end{eqnarray}
Anyone who performed a similar computation within the NPRG \cite{Papenbrock:1994kf,Codello:2013bra,Baldazzi:2020vxk} can appreciate the swiftness of this derivation.
Furthermore, equation \eqref{betaeta2loop} directly implies the $\varepsilon$-correctness of the anomalous dimension $\eta(\varepsilon) = \varepsilon^2/54 +O(\varepsilon^3)$ in $d=4-\varepsilon$, also found in the $m\to\infty$ limit of PTRG \cite{Zappala:2002nx}.
Surprisingly {\it all} the {\tt Ising} RG spectrum $\theta_i$ to order $O(\varepsilon^2)$ comes out correctly in our approach, reproducing the two-loop results $\nu = \tfrac{1}{2}+\tfrac{\varepsilon}{12}+\tfrac{7}{162}\varepsilon^2+O(\varepsilon^3)$ and $\omega = \varepsilon-\tfrac{17}{27}\varepsilon^2+O(\varepsilon^3)$ for free (details of this remarkable fact are given in the End Matter).
%


\subsection*{\color{teal}Outlook}
\vspace{-0.2cm}

The results of this Letter extend the applicability of dimensional regularization (DR) beyond the traditional limits of the $\varepsilon$-expansion, by showing that it is possible to use DR to define a natural functional RG flow -- {\it Functional Dimensional Regularization} (FDR) -- that minimizes scheme dependence (by removing altogether the need for a mass cut-off) and which furthermore shows excellent convergence properties together with critical exponents estimates that rival state-of-the-art exact RG calculations.
But what presented here is just the first part of the FDR story, since the all analysis of this paper are entirely based on one-loop master formulas, Eqs. \eqref{MasterFormulaCouplings} and \eqref{masterL}.
The key question is how to extend the sum over critical dimensions entering these formulas beyond one-loop.
A natural generalization is the following\footnote{Note that at loop order $L$ the extra dimensionality factor is $\mu^{L(d-d_c)}$.}:
\begin{equation}\label{masterL2}
\beta^{\rm FDR}(d)=\sum_L\sum_{d_{c}\in L}\mu^{L(d-d_{c})}\beta_L^{\rm DR}(d_{c})\,,
\end{equation}
where the sum over $d_c$'s is graded over the loop order $L$, and where $d_c\in L$ is the set of critical dimensions -- i.e. poles in the complex $d$--plane --  over which we sum at fixed $L$. Here one has has two basic choices: (a) decide to keep only leading order (LO) dimensions/contributions to preserve the full universality of FDR beta functions; or (b) include also scheme dependent terms that start to appear at NLO and are fully present form NNLO onwards.
A third option (c) -- partially explored in this work in relation to the computation of the anomalous dimension -- is to adopt an RG improvement point of view: one keeps only one-loop contributions, but includes all possible operators in the action -- as is usually done in functional RG.
The obvious question is if the RG improvement will make the FDR flow exact -- as in the case of the non-perturbative RG -- or not -- as in the case of the proper-time RG,  where additional correction terms are needed \cite{Litim:2001hk,Litim:2002hj,Litim:2002xm,Bonanno:2019ukb,Abel:2023ieo}.
Which is the correct way to proceed (a), (b) or (c) -- or which combination of these possibilities is the most promising -- will be the focus of future work; as well as the generalization to $O(N)$-models, non-unitary theories ({\tt Lee-Yang}) and more, to test -- with concrete examples -- the effectiveness and quality of the FDR scheme.
Finally, due to the central role that dimensional regularization has in Quantum Field Theory, the results of this Letter will have an impact well beyond Critical Phenomena, ranging from High Energy Physics to Quantum Gravity.

\subsection*{\color{teal}Acknowledgments}
\vspace{-0.2cm}

We thank D. Zappalà, G.~P. Vacca, R. Percacci and K. Falls for comments and feedback. 
The authors acknowledge financial support from the CSIC grant I+D-2022-22520220100174UD. 
A.C. also acknowledges financial support from ANII-SNI-2023-1-1013433.

\bibliography{references}

\clearpage
\onecolumngrid

\section*{End Matter}

\begin{figure}
\centering
\includegraphics[width=0.45\textwidth]{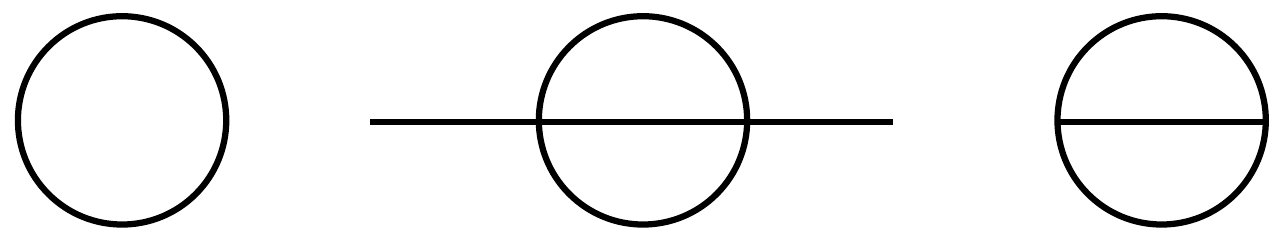}
\caption{Feynman diagrams entering the standard two-loop RG flow.}
\label{diagrams2}
\end{figure}

\begin{figure}
\centering
\includegraphics[width=0.45\textwidth]{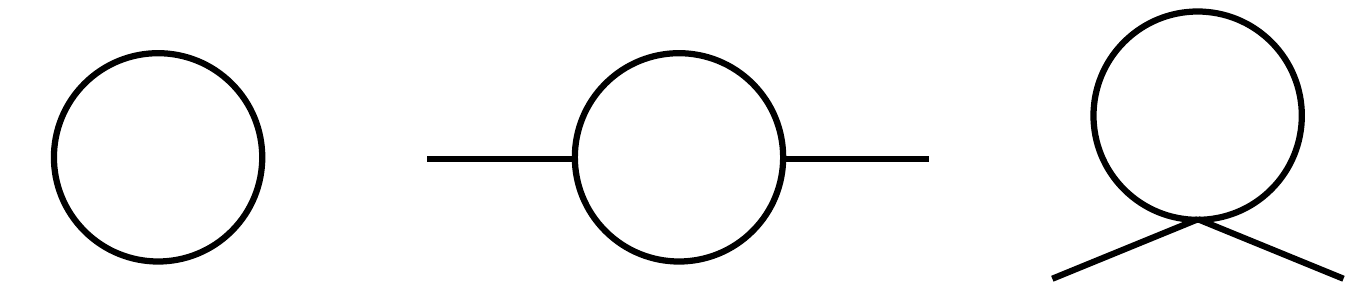}
\caption{Feynman diagrams entering the computation of the one-loop FDR flow.}
\label{diagrams1}
\end{figure}

Here we show the equivalence to order $O(\varepsilon^2)$ in $d_c=4$ between the one-loop FDR flow introduced in this Letter and the standard two-loop perturbative flow. For the latter we use the functional formalism presented in \cite{ODwyer:2007brp,Codello:2017hhh}. 
The dimension-full beta functionals in $d=4-\varepsilon$ to NLO (two-loops) are\footnote{In this End Matter contribution we drop tildes on dimension-less couplings, use the notation $v\equiv \tilde V$, $z\equiv \tilde Z $ and rescale away the usual $4\pi$ factors for clarity of the exposition.}:
\begin{eqnarray}
\beta_{v}^{\rm DR-2loop} & = & -d \, v+\tfrac{d-2+\eta}{2} \varphi  \, v'+\tfrac{1}{2}(v'')^{2}-\tfrac{1}{2}v''(v''')^{2}\nonumber\\
\beta_{z}^{\rm DR-2loop} & = & \eta\,  (1+z)-\tfrac{1}{6}(v'''')^{2} \,,\label{PTIsing}
\end{eqnarray}
where the $(v'')^2$ terms originates from the bubble diagram, $v''(v''')^2$ from the sun diagram and $(v'''')^2$ from the sunset of Fig.~\ref{diagrams2}.
The FDR flow at one-loop of Eqs. \eqref{betaV}, \eqref{betaZL1} and \eqref{DeltabetaZ}, is instead obtained from the diagrams of Fig.~\ref{diagrams1} (details of the derivation will be given in \cite{FDR}). In dimension-less form it reads
\begin{eqnarray}
\beta^{\rm FDR-1loop}_v &=& -d \, v+\tfrac{d-2+\eta}{2} \varphi  \, v'+ e^{-v''}\nonumber\\
\beta^{\rm FDR-1loop}_Z &=&
\eta \, (1+z) +\left\{- \tfrac{1}{6} (v''')^2 -z'' + \big(1+\tfrac{1}{3} v''\big)v'''z' \right\} e^{-v''}\,.
\label{FDRVZ}
\end{eqnarray}
In the beta functional $\beta^{\rm FDR-1loop}_Z$, the terms $(v''')^2$ and $v''' z'$ have their origin from the polarization of the diagrams of Fig.~\ref{diagrams1}, while the $z''$ contribution originates from the tadpole.

Beta functions are derived from the beta functionals as explained in the  main text. The DR--2loop flow \eqref{PTIsing} gives
\begin{eqnarray}
\beta_2^{\rm DR-2loop} &=& -2\lambda_2 + \lambda_4\lambda_2 -\tfrac{5}{6}\lambda_2\lambda_4^2
\nonumber\\
\beta_4^{\rm DR-2loop} &=& -\varepsilon \lambda_4 +\boxed {3\lambda_4^2 -\tfrac{17}{3}\lambda_4^3} 
+\lambda_2\lambda_6 -4\lambda_2\lambda_4\lambda_6
 \nonumber \\  \beta_6^{\rm DR-2loop} &=&  \dots \label{betaising}
\end{eqnarray}
together with the anomalous dimension $\eta=\tfrac{\lambda_{4}^{2}}{6}$.
In \eqref{betaising} we have highlighted the well known two-loop beta function of the $V(\phi) = \frac{1}{4!}\lambda_4\phi^4$ theory,
 that we have already recovered in the main text.
In DR the fixed point $\beta_i=0$ is very simple; the only non-zero coupling is the marginal one
\begin{equation}
\lambda_4^* = \tfrac{\varepsilon}{3}+\tfrac{17}{81}\varepsilon^2 +O(\varepsilon^3)\,, \label{isingfp}
\end{equation}
with all other couplings are zero. One immediately finds $\eta = \frac{\varepsilon^2}{54}+O(\varepsilon^3)$.
Finally, linear perturbations around the fixed point \eqref{isingfp} give the RG spectrum to order $O(\varepsilon^2$):
\begin{eqnarray}
\theta_2 &=&
2-\tfrac{\varepsilon}{3}-\tfrac{19}{162}\varepsilon^{2}+O(\varepsilon^3)
\nonumber\\
\theta_4 &=&
-\varepsilon+\tfrac{17}{27}\varepsilon^{2}+O(\varepsilon^3)
\nonumber\\
\theta_6 &=& -2-3 \varepsilon +\tfrac{277}{54}\varepsilon^{2}+O(\varepsilon^3) 
\nonumber\\
\theta_8 &=& -4-\tfrac{19}{3} \varepsilon +\tfrac{1282}{81}\varepsilon^{2}+O(\varepsilon^3)
\nonumber\\
 &\vdots&\label{isingspectrum}
\end{eqnarray}
For comparison with FDR we keep also the irrelevant part of the spectrum even if it can be affected by mixing with higher order operators.
%
The critical exponents are obtained as usual $\nu = 1/\theta$ and $\omega = -\theta_4$:
\begin{equation}
\nu = \tfrac{1}{2}+\tfrac{\varepsilon}{12}+\tfrac{7}{162}\varepsilon^2+O(\varepsilon^3) \qquad\qquad\qquad\omega = \varepsilon-\tfrac{17}{27}\varepsilon^2+O(\varepsilon^3)\,.
\end{equation}
We now repeat the analysis with FDR-1loop. The beta functions derived from \eqref{FDRVZ} are those of the LPA given in the main text \eqref{betadimless} with $d=4-\varepsilon$ 
%
together with those for the wave-function-renormalization couplings 
\begin{eqnarray}
\beta_{z_{2}}&=&(2-\varepsilon+2\eta)z_{2}+2\lambda_{4}^{2}F_{1}(\lambda_{2})-z_{2}^{2}F_{2}(\lambda_{2})
+z_{4}F_{2}(\lambda_{2})+z_{2}(\lambda_{4}-\lambda_{2}z_{2})F_{2}'(\lambda_{2})
+2\lambda_{4}z_{2}F_{3}(\lambda_{2})
\nonumber\\
\beta_{z_{4}}&=&...\,.
\end{eqnarray}
Here we are using the short hands,
\begin{equation}\label{F1F2F3}
F(\omega) = e^{-\omega}\qquad\qquad
F_{1}(\omega) =-\tfrac{1}{6}e^{-\omega}\qquad\qquad
F_{2}(\omega)  =-e^{-\omega}\qquad\qquad
F_3(\omega)=( 1+\tfrac{\omega}{3})e^{-\omega}\,,
\end{equation}
to track the terms of Eq. \eqref{FDRVZ}.
To order $O(\varepsilon^2)$ they exhibit the following fixed point:
\begin{eqnarray}
\lambda_{2}^* &=& \tfrac{F'(0)}{6F''(0)}\varepsilon
+\tfrac{F'(0)\left(3F''(0)^{2}-4F_{1}(0)F_{2}(0)-12F_{1}(0)F'(0)+12F'(0)F'''(0)\right)}{108F''(0)^{3}}\varepsilon^{2}+O(\varepsilon^{3})\nonumber\\
\lambda_{4}^*&=&\tfrac{\varepsilon}{3F''(0)}-\tfrac{2\left(F_{1}(0)F_{2}(0)+3F_{1}(0)F'(0)-3F'(0)F'''(0)\right)}{27F''(0)^{3}}\varepsilon^{2}+O(\varepsilon^{3})\nonumber\\
z_{2}^*&=&-\tfrac{F_{1}(0)}{9F''(0)^{2}}\varepsilon^{2}+O(\varepsilon^{3})\,,\label{fpFDR}
\end{eqnarray}
with all other couplings of order at least $O(\varepsilon^3)$. 
Note that the FP involves only $F$, $F_{1}$ and $F_{2}$.
Surprisingly when we insert \eqref{F1F2F3} into \eqref{fpFDR} we find that $\lambda_4^*$ is the same as the standard DR-2loop result \eqref{isingfp}.
The anomalous dimension that we obtain form \eqref{FDRVZ} is
\begin{equation}
\eta_{2}=\tfrac{F_{1}(0)F_{2}(0)}{9F''(0)^{2}}\varepsilon^{2}+O(\varepsilon^{3})
\end{equation}
and also agrees with the perturbative 2-loop result when we use \eqref{F1F2F3} -- as already discussed in the main text.
Finally, the RG spectrum at order $O(\varepsilon^2)$ turns out to be
\begin{eqnarray}
\theta_{2}&= & 2-\tfrac{\varepsilon}{3}+ -\tfrac{F_{1}(0)\left(F_{2}(0)-3F'(0)\right)+3F^{(3)}(0)F'(0)}{27F''(0)^{2}}\varepsilon^{2}+O(\varepsilon^{3})\nonumber\\
\theta_{4}&= &-\varepsilon+ \tfrac{2\left(F_{1}(0)\left(3F'(0)+F_{2}(0)\right)+3F^{(3)}(0)F'(0)\right)}{9F''(0)^{2}}\varepsilon^{2}+O(\varepsilon^{3})\nonumber\\
\theta_{6}&= & -2-3\varepsilon+\tfrac{15F_{1}(0)F'(0)+45F^{(3)}(0)F'(0)+7F_{1}(0)F_{2}(0)}{9F''(0)^{2}}\varepsilon^{2}+O(\varepsilon^{3})\nonumber\\
\theta_{8}&= & -4-\tfrac{19\varepsilon}{3}+\tfrac{84F_{1}(0)F'(0)+420F^{(3)}(0)F'(0)+44F_{1}(0)F_{2}(0)}{27F''(0)^{2}}\varepsilon^{2}+O(\varepsilon^{3})\label{spectrumFDR}\nonumber\\
&\vdots&
\end{eqnarray}
which still depend only on $F$, $F_{1}$ and $F_{2}$.
What is really unexpected is that \eqref{spectrumFDR}, when we use \eqref{F1F2F3}, turns out to be exactly the standard two-loop result \eqref{isingspectrum}.
This result is genuinely surprising since the two functional flows \eqref{PTIsing} and \eqref{FDRVZ} originate from different sets of diagrams at different loop orders and give rise to different sets of beta functions.
We conclude mentioning that also odd eigenvalues and the $Z$-sector -- corresponding to the operators $\phi^{i}(\partial\phi)^{2}$  -- agree to order $O(\varepsilon^2)$ with perturbative results. 
In an upcoming paper \cite{DEvsEE} we will dive deeper into these matters, and more generally test the derivative expansion of functional RGs with the $\varepsilon$-expansion.

\end{document}